\documentclass[%
 reprint,
 amsmath,
 amssymb,
 aip,
]{revtex4-2}

\usepackage{listings}
\usepackage{graphicx}
\usepackage{dcolumn}
\usepackage{bm}
\usepackage{multirow}
\usepackage{siunitx} 
\usepackage[hidelinks]{hyperref}

\usepackage[normalem]{ulem} 

\usepackage{tikz}

\begin{document}

\preprint{APS/123-QED}

\title{Martignac: Computational workflows for reproducible, traceable, and composable coarse-grained Martini simulations}

\author{Tristan Bereau}%
 \email{bereau@uni-heidelberg.de}
\affiliation{%
 Institute for Theoretical Physics, Heidelberg University, 69120 Heidelberg, Germany
}%
\author{Luis J.~Walter}
\affiliation{%
Institute for Theoretical Physics, Heidelberg University, 69120 Heidelberg, Germany
}
\author{Joseph F.~Rudzinski}
\affiliation{Physics Department and CSMB Adlershof, Humboldt-Universität zu Berlin, 12489 Berlin, Germany}

\date{\today}

\begin{abstract}
Despite their wide use and far-reaching implications, molecular dynamics (MD) simulations suffer from a lack of both traceability and reproducibility. We introduce Martignac: computational workflows for the coarse-grained (CG) Martini force field. Martignac describes Martini CG MD simulations as an acyclic directed graph, providing the entire history of a simulation---from system preparation to property calculations. Martignac connects to NOMAD, such that all simulation data generated are automatically normalized and stored according to the FAIR principles. We present several prototypical Martini workflows, including system generation of simple liquids and bilayers, as well as free-energy calculations for solute solvation in homogeneous liquids and drug permeation in lipid bilayers. By connecting to the NOMAD database to automatically pull existing simulations and push any new simulation generated, Martignac contributes to improving the sustainability and reproducibility of molecular simulations.
\end{abstract}

\maketitle

\section{Introduction}

\begin{figure}[htbp]
    \centering
    \includegraphics[width=0.7\linewidth]{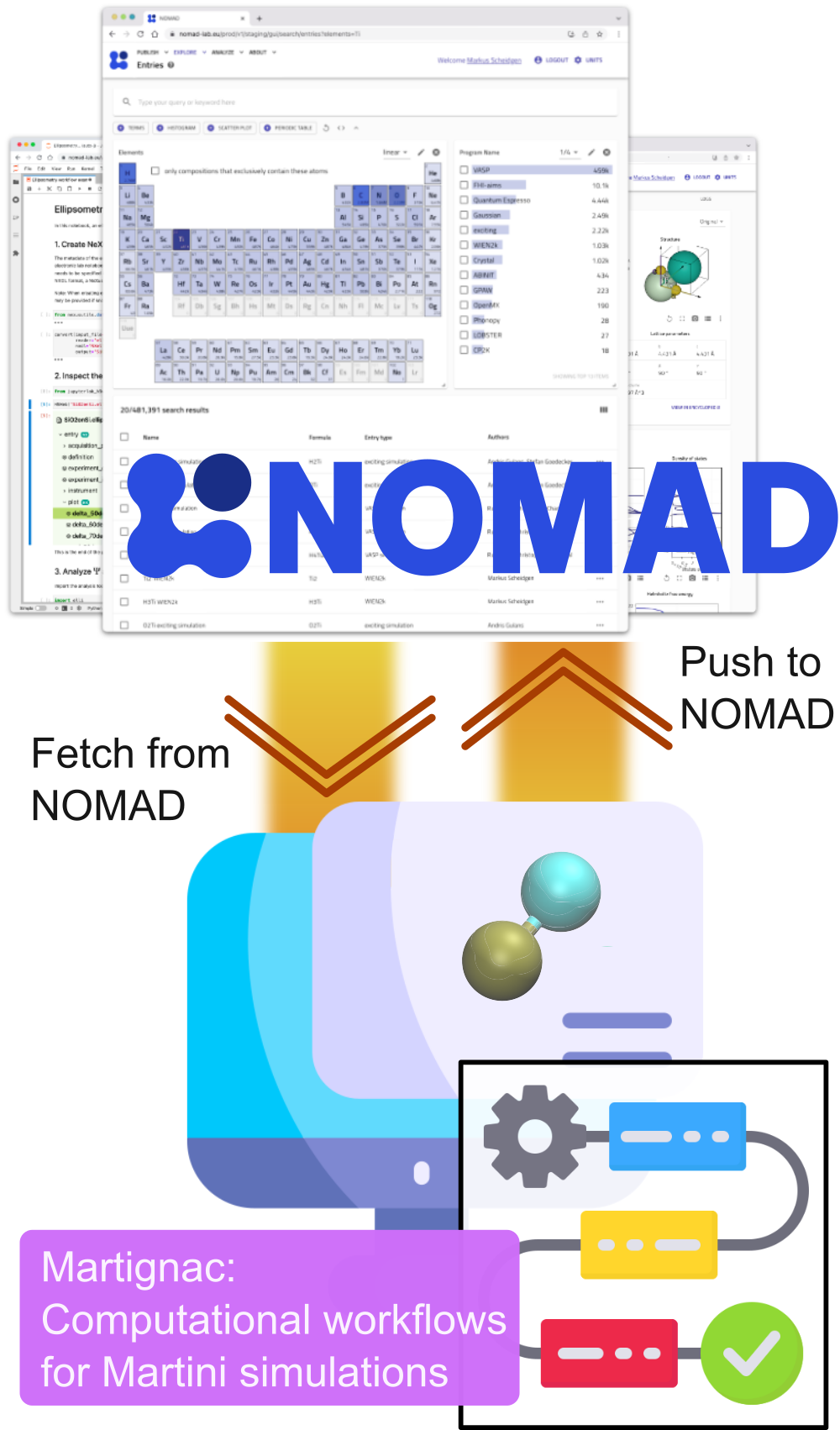}
    \caption{Martignac implements Martini coarse-grained simulations as computational workflows. The library interacts with the NOMAD web server to automatically fetch any existing simulation, and push any new contribution.}
    \label{fig:nomad-intro}
\end{figure}

Despite ever-increasing attention and community efforts for the last half century, molecular dynamics (MD) simulations remain poorly shared, deficiently reproducible, and often devoid of history or traceability.  The sharing of MD data is made complex due to the fragmentation of hardware, software code, force fields, and simply the sheer diversity of systems of interest.\cite{abraham2019sharing} Efforts in this direction include the BioExcel Building Blocks (BioBB) library,\cite{andrio2019bioexcel} the COVID-19 Molecular Structure and Therapeutics Hub,\cite{covid_molssi} a general index of MD-simulation repositories found online (MDverse),\cite{mdverse} and the Simulation Foundry.\cite{gygli2020simulation} Yet, a recent document reiterated specific needs for the community, including persistent, indexed, and open access to MD data, metadata annotation, application programming interfaces (APIs) for data exchange, and comprehensive provenance information (i.e., history of the simulation).\cite{amaro2024need} Here we propose a concrete end-to-end solution for the popular coarse-grained (CG) Martini biomolecular force field.\cite{marrink2007martini, souza2021martini} We introduce \emph{Martignac}, a workflow manager that automatically connects to an online database, avoids redundant calculations by downloading existing entries, runs missing simulations, and subsequently uploads them to enrich the database (Fig.~\ref{fig:nomad-intro}). Martignac offers a traceable, composable, and reproducible framework for general-purpose and high-throughput CG Martini simulations.

To enable reproducibility and provenance information, computer simulations need a strict, specialized, and systematic \emph{workflow}: a management system designed to orchestrate activities and organize resources into processes.  Scientific workflows typically compose individual units of calculations as a directed acyclic graph (DAG). They have rapidly become an essential ingredient to capture provenance information in materials modeling (e.g., see AiiDA workflows\cite{uhrin2021workflows}). Martignac makes integral use of computational workflows. We will see that not only does this offer reproducible and traceable MD simulations, it also enables the \emph{composability} of several MD workflows together, thus avoiding redundant calculations upon compound screening.

Martignac implements worklows via \texttt{signac-flow}, a Python library that manages and automates workflows in computational research.\cite{adorf2018simple, ramasubramani2018signac} It organizes, executes, and monitors data processing pipelines, making it easier to handle complex and large-scale simulations and analyses. The framework has been applied to several projects, including the assembly of colloidal diamond,\cite{zhou2022route} photonic crystals,\cite{cersonsky2021diversity} and lubricating monolayer films.\cite{summers2020mosdef, quach2022high} The last application mentioned is the outcome of a larger consortium called MoSDeF, which offers a set of Python tools to help initialize, assign force-field parameters, and support screening of soft-matter systems.\cite{mosdef} Several of the authors of the last study have highlighted \texttt{signac-flow} as an essential tool to improve the reproducibility of molecular simulations, where they propose a set of principles to create Transparent, Reproducible, Usable by others, and Extensible (TRUE) molecular simulations.\cite{thompson2020towards} Though not mentioned in the article, the TRUE principles overlap significantly with the much more widespread FAIR data principles: Findability, Accessibility, Interoperability, and Reuse of digital assets.\cite{wilkinson2016fair} 

The aspiration to work with FAIR data has seen rapid and intense developments in many areas of science.\cite{tanhua2019ocean, jeliazkova2021towards, jacobsson2022open} In materials science, NOMAD\cite{nomad_home} is one of the leading efforts in building FAIR databases.\cite{draxl2018nomad, draxl2019nomad, scheffler2022fair} Originally built as a repository for {\it ab initio} calculations, NOMAD has been recently transformed into a versatile research data management platform for a wide variety of materials science data.\cite{scheidgen2023nomad} Specifically relevant for this work, the openly-available NOMAD web-based platform provides the following functionalities: ($i$) automated detection and parsing of raw molecular simulation files from GROMACS, ($ii$) custom workflows that allow connections between independently run simulations and analysis stored in the database, and ($iii$) a full suite of API commands, enabling scriptable communication with the database. Martignac leverages these functionalities to not only facilitate transparent storage of the executed simulations and workflows but also to improve efficiency and prevent redundancy, i.e., to provide a comprehensive FAIR data management solution.

Martignac places a specific emphasis on high-throughput screening (HTS) applications.  The Martini model has been an invaluable model for HTS applications due to its top-down parametrization: its building-block approach of CG bead types effectively reduces the size of chemical compound space.\cite{kanekal2019resolution, bereau2021computational} The use of Martini for HTS has enabled a number of applications, including protein-ligand binding,\cite{souza2020protein} the extensive screening of drug-membrane permeation for 0.5 million compounds,\cite{menichetti2019drug} the potentials of mean force (PMFs) of all Martini dimers inserted in six phospholipid bilayers,\cite{hoffmann2020molecular} the identification of driving forces for generic anaesthetics,\cite{centi2020inserting} and the molecular discovery of a molecular probe selective to cardiolipin.\cite{mohr2022data}

Martignac implements a select list of computational workflows that is expected to be of general interest to the broader CG Martini community. We distinguish two categories of workflows: ($i$) system generation and ($ii$) free-energy calculation. The systems that Martignac can generate consist of: a solute in the gas (i.e., a molecule in an empty simulation box); a solvent (i.e., homogeneous fluid); a solute inserted in a solvent; and a phospholipid bilayer. The free-energy-calculation workflows comprise the solvation of a solute in a solvent, and the potential of mean force of a small molecule in a phospholipid bilayer. The reader may already foretell the workflow composability at play: for instance, calculating the solvation free energy of a solute requires to both generate the solute and the solvent independently, then combine them to generate the system, and finally run the free-energy calculations.

The present article first describes the NOMAD database in Sec.~\ref{sec:nomad}, followed by Martignac's workflow methodology in Sec~\ref{sec:workflow-method}, including the contents of the worfklows and how they connect to NOMAD. In Sec.~\ref{sec:simulation_methods} we summarize some of the MD simulation methods and parameters. Sec.~\ref{sec:results} highlights a number of applications made possible by Martignac: how the content of the DAG generated by Martignac is translated to NOMAD, the composability of workflows, a reproducibility calculation of oil/water transfer free energies, and finally another reproducibility calculation for drug-membrane PMFs. We conclude with a number of final remarks and outlook in Sec.~\ref{sec:conclusion}.

\section{NOMAD functionalities}
\label{sec:nomad}

This section on the NOMAD database is only a short summary of the NOMAD documentation,\cite{nomad_doc} with a focus on aspects useful to Martignac.

\subsection{Processing and organization}

NOMAD ingests the raw input and output files from standard simulation software by first identifying a representative file (e.g., the \texttt{log} file in the case of GROMACS) and then employing a parser code to extract relevant (meta)data from the associated files to that simulation. The (meta)data are stored within a structured schema---the NOMAD Metainfo---to provide context for each quantity, enabling interoperability and comparison between simulation software. The compilation of all (meta)data obtained from this processing forms an \emph{entry}---the fundamental unit of storage within the NOMAD database---including simulation input/output, author information, and additional general overarching metadata (e.g., references or comments). In addition, a NOMAD entry offers unique identifiers: both an \texttt{entry\_id} to manage unpublished data and also a DOI once published.

NOMAD entries can be organized hierarchically into \emph{uploads}, \emph{workflows}, and \emph{datasets}. Since the parsing execution is dependent on automated identification of representative files, users are free to arbitrarily group simulations together upon upload. In this case, multiple entries will be created with the corresponding simulation data. An additional unique identifier, \texttt{upload\_id}, will be provided for this group of entries. Although the grouping of entries into an upload is not necessarily scientifically meaningful, it is practically useful for submitting batches of files from multiple simulations to NOMAD. Concretely, Martignac utilizes uploads to group all $\lambda$ coupling points of a thermodynamic-integration calculation. This is particularly convenient since NOMAD retains the original directory structure when storing all the raw and processed data.

NOMAD offers flexibility in the construction of workflows. First, a molecular dynamics simulation is a workflow in itself. The (meta)data for this standard workflow are automatically stored during processing and entail all relevant aspects: MD runtime parameters and ensemble-averaged or time-dependent observables. Furthermore, NOMAD also allows the creation of custom workflows, which are completely general directed graphs, allowing users to link NOMAD entries with one another in order to provide the \emph{provenance} of the simulation data. Custom workflows are contained within their own entries and, thus, have their own set of unique identifiers. To create a custom workflow, the user is required to upload a workflow \texttt{yaml} file describing the inputs and outputs of each entry within the workflow, with respect to sections of the NOMAD Metainfo schema. Martignac automatically creates this file for its workflows, without the user being required to understand any details of the NOMAD schema.

At the highest level, NOMAD groups entries with the use of \emph{datasets}. A NOMAD dataset allows the user to group a large number of entries, without any specification of links between individual entries. A DOI is also generated when a dataset is published, providing a convenient route for referencing all data used for a particular investigation within a publication.

\subsection{Programmatic access, query, and interaction}

In addition to its GUI interface, NOMAD supports scriptable access to its database and functionalities through an extensive application programming interface (API).\cite{scheidgen2023nomad} A REST API queries the server by a combination of GET and POST \texttt{requests}. Martignac uses this API through a variety of python functions that lower the barrier for use by conveniently combining multiple API calls into a single routine and perform validation tests. In particular Martignac uses Marshmallow schemas to validate the input data that is received, as well as deserialize the input data to Python objects.\cite{marshmallow} Read-only requests of publicly available entries can be made without NOMAD credentials. Otherwise, a NOMAD username and password is required to authenticate the API request.

\section{Martignac workflow methodology}
\label{sec:workflow-method}

The present section focuses on the conceptual ideas behind Martignac. For a description of the Python library, we refer the reader to the online documentation.\cite{martignac_doc}

Workflows are built as directed acyclic graphs (DAGs). DAGs consist of nodes and single-directional edges. In our context, nodes are computational \emph{operations}, e.g., generating a molecular configuration or running an MD simulation. These operations are interlinked---one cannot analyze a trajectory that has not yet been simulated---leading to a required ordering of the operations via directional connections, i.e., the DAG's edges. 

The Martignac implementation builds on the flexible \texttt{signac-flow} library.\cite{adorf2018simple, ramasubramani2018signac} \texttt{signac-flow} systematically manages and distributes a set of operations (e.g., MD simulations) repeated across many systems. This makes \texttt{signac-flow} appealing for high-throughput screening, where consistency is a paramount requirement. Each system studied is called a \emph{state point}, which \texttt{signac-flow} associates to a unique and persistent job identifier (ID). Each job is thereby a collection of interdependent operations, each implemented as a Python function. In addition, the workflow requires label functions, which determine whether the operation has already been run. Finally, interdependences between operations (i.e., the DAG's edges) are specified by pre- and post-conditions on other label functions. 

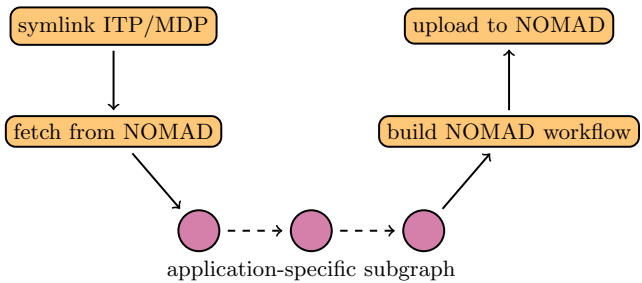
\begin{figure}[htbp]
    \centering
    \vspace{0.5cm}
    \resizebox{\columnwidth}{!}{
        \begin{tikzpicture}[node distance={20mm}, thick, main/.style = {draw, rectangle, rounded corners}]
        \node[main,fill={rgb:orange,1;yellow,2;pink,5}] (1) {symlink ITP/MDP}; 
        \node[main,fill={rgb:orange,1;yellow,2;pink,5}] (2) [below of=1, yshift=0.5cm] {fetch from NOMAD};
        \node[main,circle,fill={rgb:red,1;blue,1;pink,4}] (3a) [below right of=2, xshift=-0.2cm] {\quad\quad}; 
        \node[main,circle,fill={rgb:red,1;blue,1;pink,4},label=below:application-specific subgraph] (3b) [right of=3a, xshift=-0.4cm] {\quad\quad}; 
        \node[main,circle,fill={rgb:red,1;blue,1;pink,4}] (3c) [right of=3b, xshift=-0.4cm] {\quad\quad}; 
        \node[main,fill={rgb:orange,1;yellow,2;pink,5}] (4) [above right of=3c, xshift=-0.2cm] {build NOMAD workflow};
        \node[main,fill={rgb:orange,1;yellow,2;pink,5}] (5) [above of=4, yshift=-0.5cm] {upload to NOMAD};
        \path[->,shorten <= 2pt, shorten >= 2pt] (1) edge node[sloped,below] {} (2);
        \path[->,shorten <= 3pt, shorten >= 3pt] (2) edge node[sloped,below] {} (3a);
        \path[->,shorten <= 3pt, shorten >= 3pt] (3a) edge [dashed] node[sloped,below] {} (3b);
        \path[->,shorten <= 3pt, shorten >= 3pt] (3b) edge [dashed] node[sloped,below] {} (3c);
        \path[->,shorten <= 3pt, shorten >= 3pt] (3c) edge node[sloped,below] {} (4);
        \path[->,shorten <= 2pt, shorten >= 2pt] (4) edge node[sloped,below] {} (5);
        \end{tikzpicture}
    }
    \caption{The Martignac workflow structure is split in two subgraphs: ($i$) generic operations common to all simulation workflows (yellow); ($ii$) domain-application specific operations (see Fig.~\ref{fig:workflows}). \label{fig:generic_dag}}
\end{figure}

Martignac workflows target various domain applications (e.g., generating a solvent or calculating the free energy of a solute in a phospholipid membrane). Because these domain applications share a number of operations, we separate workflows into two components:
\begin{enumerate}
    \item Generic operations that we apply irrespective of the domain application;
    \item Domain-application specific operations;
\end{enumerate}
as shown in Fig.~\ref{fig:generic_dag}.

\subsection{Generic subgraph}

The generic operations consist of the following:
\begin{itemize}
    \item \textbf{symlink ITP/MDP}: Force-field-definition files are extremely redundant: the same sets of \texttt{*.itp} files are necessary to run any Martini simulation. Similarly, high-throughput workflows will typically work with identical \texttt{*.mdp} input-parameter files to obtain systematic simulations. A systematic upload of these files creates an unnecessary burden on storage requirements. To this end, Martignac makes use of symbolic links to mitigate local storage footprint and avoids uploading said files to the server. However, reproducibility remains: standard Martini force-field files are available online.\cite{martini_ff} While input \texttt{*.mdp} parameter files are not saved, Martignac does store the (equivalent) output \texttt{*.mdp} files generated by the GROMACS preprocessor, \texttt{grompp}.
    
    \item \textbf{fetch from NOMAD}: Martignac queries the user's NOMAD dataset to look for an existing simulation already stored online. Martignac checks whether the simulation queried \emph{exactly} corresponds to the one to be attempted. Comparison is made on the basis of ($i$) the workflow Python-class name, ($ii$) the \texttt{signac-flow} job ID, and ($iii$) a hash of all input \texttt{*.mdp} files involved. The information is contained as a JSON-formatted comment of the simulation upload, which is automatically generated and pushed with any upload (see below). For instance, Fig.~\ref{fig:job_comment} is a comment for a simulation upload of a single solute generation. This check ensures integrity beyond the mere desired workflow and chemistry, but also in the exact input files used. 
    \item \textbf{build NOMAD workflow}: The chain of operations implemented in Martignac form a DAG. Said DAG is converted and serialized into a NOMAD-compatible workflow \texttt{yaml} file, for subsequent simulation upload.
    \item \textbf{upload to NOMAD}: All files generated during a Martignac workflow (except for the \texttt{*.itp} Martini definition files and input \texttt{*.mdp}s) are zipped together with the \texttt{yaml} NOMAD workflow file and uploaded to the NOMAD webserver via an API \texttt{POST} request. A comment is attached to every upload containing a JSON-formatted string containing identifiable information about the content of the job. 
\end{itemize}
The generic operations are arranged as shown in Fig.~\ref{fig:generic_dag}: a linear chain of operations with the application-specific subgraph in the middle.

\begin{figure}[htbp]
        \begin{lstlisting}[language=Python]
{'job_id': '3e793a7b2a1e83233c40458fddf958ab',
 'itp_files': 'a52590b1d87d122ba1e376b83c3d6bee', 
 'mdp_files': '2d7a9e52d14d23e0dfb97192d75a3463',
 'state_point': {
    'solute_name': 'P5', 'type': 'solute'
 }, 
  'workflow_name': 'SoluteGenFlow'}
    \end{lstlisting}
    \caption{Comment of an example job upload. The respective keys correspond to the \texttt{signac-flow} job ID, the hashes of the collection of \texttt{*.itp} and \texttt{*.mdp} files, respectively, state-point dependent information, including the solute name and type, and finally the workflow Python-class name. \label{fig:job_comment}}
\end{figure}

\subsection{Application-specific subgraphs}

Here we shortly describe the directed acyclic subgraphs that are application specific, and contained within the larger Martignac DAG (see Fig.~\ref{fig:generic_dag}). This implies that all application-specific subgraphs described below are both preceded and followed by the node operations described in the generic subgraph above.

\begin{figure*}[htbp]
    \centering
    \includegraphics[width=0.9\linewidth]{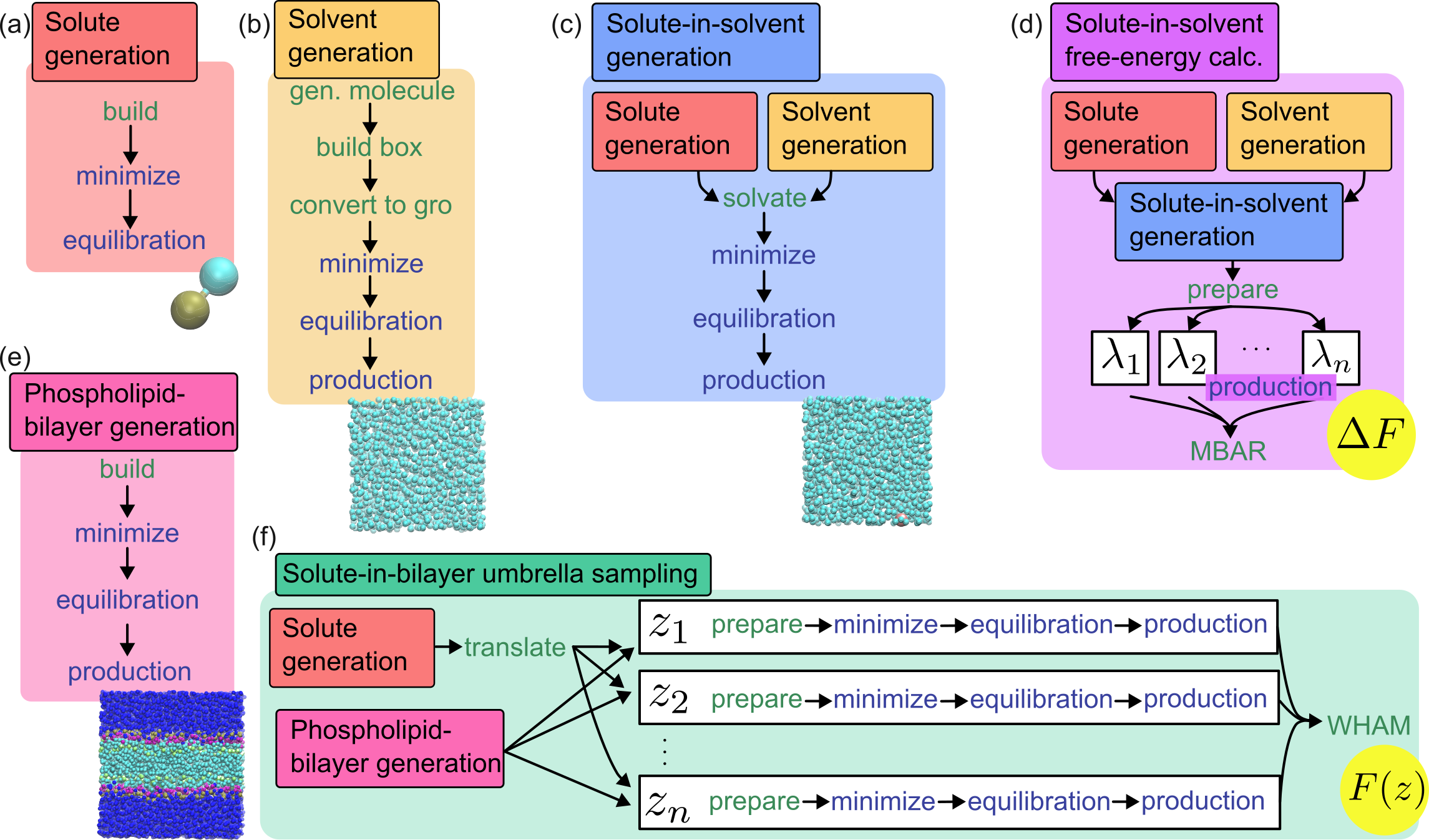}
    \caption{Martignac workflows. System generation: (a) Solute; (b) solvent; (c) solute in solvent; (e) phospholipid bilayer. Free-energy calculations: (d) solute-in-solvent free energy; (f) solute-in-bilayer potential of mean force. Blue and green operations distinguish MD simulations from the others, respectively. The composability of workflows is highlighted by entire graphs being part of others. Aggregation of $\lambda$ states or umbrella collective variables, $z$, recycles common operations and makes use of all relevant information for analysis. The outcome of each workflow is illustrated: either a generated system or a free energy.}
    \label{fig:workflows}
\end{figure*}

\subsubsection{Solute generation}
\label{sec:solute_gen}

Generating a solute in the gas phase (i.e., an empty box, Fig.~\ref{fig:workflows}a) consists of three steps
\begin{enumerate}
    \item \textbf{build}: From the name of the solute molecule, generate a structure, particle-definition, and topology files, \texttt{gro}, \texttt{itp}, and \texttt{top}, respectively.
    \item \textbf{minimize}: Energy minimization.
    \item \textbf{equilibrate}: Equilibration MD simulation.
\end{enumerate}
The only state-point parameter is the name of the solute.

\subsubsection{Solvent generation}
\label{sec:solvent_gen}

We consider the generation of a homogeneous liquid that fills the simulation box (Fig.~\ref{fig:workflows}b)
\begin{enumerate}
    \item \textbf{generate solvent molecule}: Generate a single solvent molecule. This step is analogous to the \textbf{build} part of the solute-generation workflow.
    \item \textbf{build solvent box}: build a box of solvent molecules by means of the PACKMOL program.\cite{martinez2009packmol}
    \item \textbf{convert box to gro}: The preceding PACKMOL operation yields a \texttt{pdb} file. The present operation simply converts the \texttt{pdb} to a \texttt{gro} structure file.
    \item \textbf{minimize}: Energy minimization.
    \item \textbf{equilibrate}: Equilibration MD simulation. For instance, this could be run at constant pressure with a forgiving barostat, such as Berendsen or C-rescale.
    \item \textbf{production}: Production MD simulation.
\end{enumerate}
The only state-point parameter is the name of the solvent molecule.

\subsubsection{Solute-in-solvent generation}
\label{sec:solute_solvent_gen}

The solute-in-solvent generation (Fig.~\ref{fig:workflows}c) makes use of the \emph{composability} of the Martignac workflows: it first generates the solute and the solvent using the above-mentioned workflows, and subsequently joins them to yield the desired system. As such the DAG is not linear, but contains branches to link the individual components together.
\begin{enumerate}
    \item \textbf{generate solute}: fetch or run the solute generation workflow (Sec.~\ref{sec:solute_gen}).
    \item \textbf{generate solvent}: fetch or run the solvent generation workflow (Sec.~\ref{sec:solvent_gen}).
    \item \textbf{solvate}: solvate the solute using the GROMACS \texttt{gmx solvate} program.
    \item \textbf{minimize}: Energy minimization.
    \item \textbf{equilibrate}: Equilibration MD simulation.
    \item \textbf{production}: Production MD simulation.
\end{enumerate}
The state-point parameters are the solute and solvent names.

\subsubsection{Solute-in-solvent alchemical transformation}
\label{sec:solute_solvent_alchemical}

Composability is leveraged once again here (Fig.~\ref{fig:workflows}d). We make use of the solute-in-solvent system generated in the last workflow. Free-energy calculations are employed to compute the free energy of coupling the solute in the solvent. We make use of thermodynamic integration, where a series of Hamiltonians interpolating between the two end states, denoted by the parameter $\lambda \in [0,1]$, increasingly couple the solute in the simulation box. As such the DAG needs to be run not only once for the system of interest, but $n$ times, indicative of the number of interpolating Hamiltonians. The present workflow DAG is thereby \emph{nonlinear}: it will run MD simulations for each $\lambda$ state in parallel. However, both the system initialization and the final free-energy calculation ought to occur only once. This is illustrated in Fig. \ref{fig:workflows} (d). Concretely, this is implemented by means of an \emph{aggregator} function decorator.
\begin{enumerate}
    \item \textbf{prepare system}: Fetch or run the solute generation, solvent generation, and solute-in-solvent generation workflows. (Aggregated operation.)
    \item \textbf{production}: Production MD simulation at a specific $\lambda$ value, additionally evaluating and storing the energy from all interpolating Hamiltonians, $U_\lambda$, for later use in the free-energy calculations.
    \item \textbf{compute free energy}: Compute the free energy by means of the multi-Bennett acceptance ratio (MBAR)\cite{shirts2008statistically} via the \texttt{alchemlyb} library.\cite{oliver_beckstein_2024_12692737} (Aggregated operation.)
\end{enumerate}
The state-point parameters are the solute name, solvent name, and Hamiltonian-coupling $\lambda$ value.

\subsubsection{Phospholipid-bilayer generation}
\label{sec:bilayer_gen}

Here we consider the generation of a phospholipid bilayer (Fig.~\ref{fig:workflows}e). The implementation not only allows for a variety of single-composition (Martini-supported) lipid bilayers, it also supports the generation of lipid mixtures.
\begin{enumerate}
    \item \textbf{generate initial bilayer}: Generation of an initial phospholipid-bilayer structure by means of the INSANE tool.\cite{wassenaar2015computational}
    \item \textbf{minimize} Energy minimization.
    \item \textbf{equilibrate} Equilibration MD simulation.
    \item \textbf{production} Production MD simulation.
\end{enumerate}
The relevant state-point parameters are the name and fractional composition of each phospholipid name.

\subsubsection{Solute-in-bilayer umbrella sampling}
\label{sec:bilayer_umbrella}

We consider the thermodynamics of insertion of a solute molecule in a phospholipid bilayer (Fig.~\ref{fig:workflows}f). We compute the potential of mean force (PMF) by means of umbrella sampling.\cite{torrie1977nonphysical} We consider the PMF of insertion against a typical collective variable: the depth normal to the bilayer, $z$. This last workflow is again a combination of composability and state-point aggregation: Composability of the solute generation (Sec.~\ref{sec:solute_gen}) with phospholipid-bilayer generation (Sec.~\ref{sec:bilayer_gen}); and state-point aggregation when collecting umbrella-sampling restraints placed at various intervals of the collective variable, $z$. 
\begin{enumerate}
    \item \textbf{generate solute}: fetch or run the solute generation workflow (Sec.~\ref{sec:solute_gen}). (Aggregated operation.)
    \item \textbf{translate solute}: move the solute to the origin of the simulation box. (Aggregated operation.)
    \item \textbf{generate bilayer}: fetch or run the phospholipid-bilayer generation workflow (Sec.~\ref{sec:bilayer_gen}). (Aggregated operation.)
    \item \textbf{insert solute in box}: use PACKMOL to place the solute in the bilayer box.
    \item \textbf{convert box to gro}: Convert PACKMOL's output \texttt{pdb} file to \texttt{gro} format.
    \item \textbf{update topology file}: Combine the topology files of the solute and bilayer systems.
    \item \textbf{minimize} energy minimization.
    \item \textbf{equilibrate} MD-based equilibration simulation.
    \item \textbf{production} Production MD simulation.
    \item \textbf{compute WHAM}: Use GROMACS' implementation of the weighted histogram analysis method (WHAM) to compute the PMF.\cite{kumar1992weighted} (Aggregated function.)
    \item \textbf{analyze WHAM}: Convert and store the GROMACS WHAM output \texttt{xvg} files to \texttt{numpy} arrays.
\end{enumerate}

\section{Simulation methods}
\label{sec:simulation_methods}

Molecular dynamics (MD) simulations were performed with GROMACS 2023.1.\cite{abraham2015gromacs} Unless specified, we relied on the Martini 3 force-field parameters\cite{souza2021martini} with an integration time step of $\delta t = 0.02~\tau$, where $\tau$ is the model's natural unit of time. Simulations targeted an $NPT$ ensemble: constant number of particles, pressure ($P=1$~bar), and temperature ($T=298$~K). The latter was controlled by means of a stochastic velocity-rescaling thermostat.\cite{bussi2007canonical} Equilibration MD simulations typically made use of the Berendsen or C-rescale barostats, while production simulations relied on the more accurate, but also more sensitive, Parrinello-Rahman barostat.\cite{parrinello1981polymorphic}

To generate solvent boxes, we used the PACKMOL program,\cite{martinez2009packmol} and INSANE was used to generate phospholipid bilayers.\cite{wassenaar2015computational} Various tools of the GROMACS suite were used to generate \texttt{gro} structures and topology files, solvate a solute, or run the weighted histogram analysis method (WHAM). Alchemical free energies were calculated by means of the multi-Bennett acceptance ratio (MBAR)\cite{shirts2008statistically} via the \texttt{alchemlyb} library.\cite{oliver_beckstein_2024_12692737}

Because of variations in the exact protocol used in the various workflows, we refer the reader to the Martignac implementation or published NOMAD entries for more detailed information. In particular, the full set of parsed simulation input parameters can be easily browsed via NOMAD's MetaInfo viewer, found under the ``Data'' tab of each entry page.

\section{Results}
\label{sec:results}

This section highlights a number of features enabled by Martignac's computational-workflow design. To accompany the results, we systematically refer to the hyperlinked NOMAD upload ID for easy access to each computational workflow and underlying simulations. We also provide a graphical user interface to the NOMAD uploaded Martignac uploads in a web-based app on Streamlit.\cite{streamlit} The Streamlit app dynamically queries NOMAD, and features application-specific properties, such as the underlying DAG, free energies, and potentials of mean force.

\subsection{The Martignac directed graph translates to NOMAD metadata}

\begin{figure*}[htbp]
    \centering
    \includegraphics[width=0.85\linewidth]{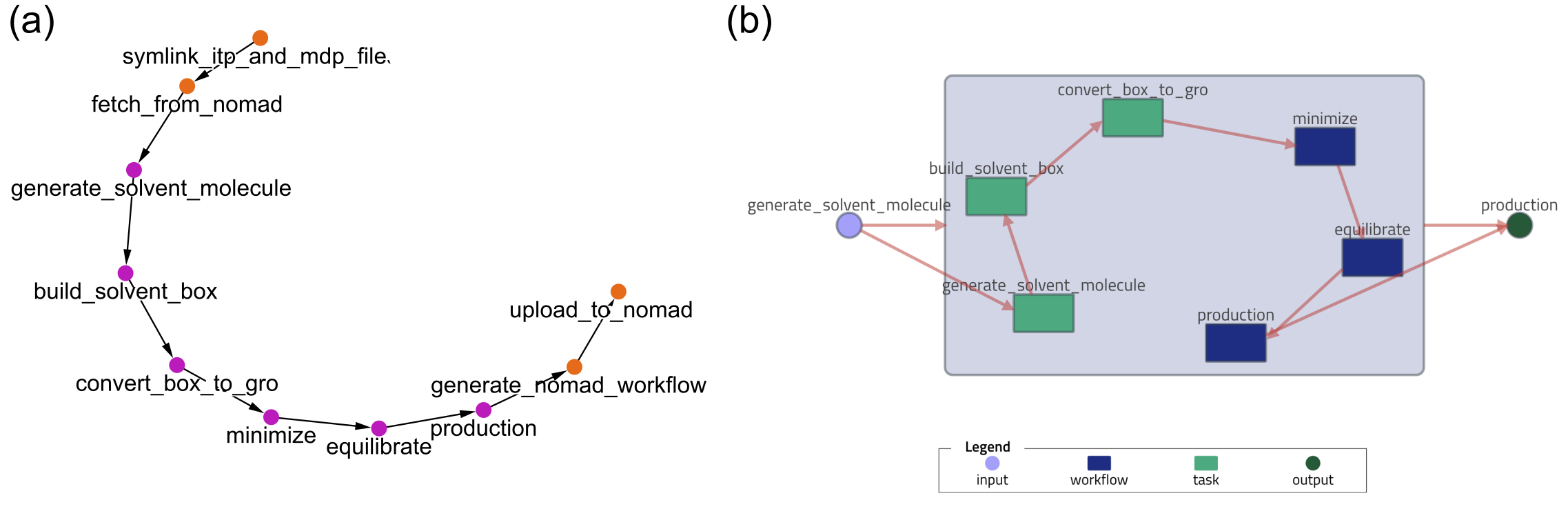}
    \caption{Workflow graph for solvent generation. (a) Workflow generated from Martignac. Color coding follows Fig.~\ref{fig:generic_dag}. (b) Workflow generated from NOMAD. Green and dark-blue rectangles display operations, and distinguish those that involve an MD simulation.}
    \label{fig:solvent-workflow}
\end{figure*}

As a first example of the interaction between Martignac and NOMAD, we consider the generation of a box of hexadecane molecules---one of the standard Martini solvents. Fig.~\ref{fig:solvent-workflow} (a) shows a DAG that is automatically generated by reading Martignac's set of operations and pre/post-conditions. The DAG contains all generic and application-specific operations. Though this DAG is linear, others in this work have non-trivial connectivity, owing to loading several other workflows as part of the early system setup, or aggregation of simulations for free-energy calculations. 

In comparison to the Martignac DAG, we also show in Fig.~\ref{fig:solvent-workflow} (b) an illustration of the workflow that is generated and interpreted by NOMAD. NOMAD correctly identifies all operations, and even distinguishes operations that consist of MD simulations from the others. NOMAD's correct visualization of the workflow validates the programmatic transfer of the DAG into simulation metadata. An example can be found for the (hyperlinked) NOMAD upload ID \href{https://nomad-lab.eu/prod/v1/staging/gui/user/uploads/upload/id/uzssztc-SrGcz49GuSVlqQ}{\texttt{uzssztc-SrGcz49GuSVlqQ}}. 

\subsection{Workflow composability}

Avoiding unnecessary redundancies is an important feature of high-throughput calculations, because they can save significant compute and storage resources. For instance, the screening of the insertion of a solute in a solvent involves compounding combinatorics: each solute against each solvent. We avoid generating the same system twice by enforcing composability. Several workflows start with the generation of the elementary parts, e.g., solute and solvent in a solute-in-solvent system, or solute and lipid bilayer in drug permeation. Each elementary step relies on a ``fetch-or-run'' mechanism: we first check whether the system has been previously run by means of an API call. If so, we download it, otherwise, we run the system. If downloaded, Martignac locally stores the NOMAD upload ID of the elementary workflow. The elementary workflow is included in the final NOMAD workflow by referencing the elementary upload ID. This referencing of existing workflows enables a hierarchical structure and reusability of simulations. We check that when running a high-throughput calculation of solute-in-a-solvent generation, a given chemistry for a solute is ever calculated and stored only once. Case in point: the alchemical calculation for the solute bead P6 in a homogeneous liquid of hexadecane is stored in a single upload with (hyperlinked) ID \href{https://nomad-lab.eu/prod/v1/staging/gui/user/uploads/upload/id/PCQSjL2wQsCptzZhHal96Q}{\texttt{PCQSjL2wQsCptzZhHal96Q}}. Every $\lambda$-point simulation relies on the same solute system generation \href{https://nomad-lab.eu/prod/v1/staging/gui/user/uploads/upload/id/oT0F5qP9RY6AFDiDQrJ2ug}{\texttt{oT0F5qP9RY6AFDiDQrJ2ug}}, which is built from a single solute generation, \href{https://nomad-lab.eu/prod/v1/staging/gui/user/uploads/upload/id/7YU8feV_SQ6h6M7aIUCdQg}{\texttt{7YU8feV\_SQ6h6M7aIUCdQg}}, and a single solvent generation, \href{https://nomad-lab.eu/prod/v1/staging/gui/user/uploads/upload/id/uzssztc-SrGcz49GuSVlqQ}{\texttt{uzssztc-SrGcz49GuSVlqQ}}.

\subsection{Reproducibility of oil/water transfer free energies}

Martignac facilitates reproducibility by the systematic nature of its computational workflows. As a first example, we focus on oil/water transfer free energies. The recent Martini 3 force field provides an extensive reference of free-energy calculations as supporting information.\cite{souza2021martini} Parts of these reference thermodynamic calculations include oil/water transfer free energies for the majority of CG beads defined by the Martini model.  Here we reproduce a subset of these calculations by means of the \emph{Solute-in-solvent alchemical transformation} workflow (Fig.~\ref{fig:workflows}d). Because the workflow incorporates not only system generation and MD simulations, but also the calculations of the free energies themselves, these are straightforward to store as metadata in NOMAD. As such, we directly query the free energies from NOMAD to fetch easily- and permanently-available thermodynamic properties. 

\begin{table}[h!]
\centering
\begin{tabular}{r|r||rr}
Solute & Solvent & Martignac & Reference \\
\hline
\multirow{5}{*}{P6} & HD $\rightarrow$ W & -27.45 & -27.20 \\
 & CLF $\rightarrow$ W & -11.98 & -11.90 \\
 & ETH $\rightarrow$ W & -10.96 & -11.20 \\
 & CHEX $\rightarrow$ W & -18.86 & -19.00 \\
 & W & 17.98 & 18.00 \\
\end{tabular}
\caption{\label{tab:transfer-free} Reproducibility of solute-in-solvent free energies against the Martini 3 publication.\cite{souza2021martini} Solvents with and without a right-pointing arrow denote transfer and hydration free energies, respectively. All free energies in units of kJ/mol.}
\end{table}

Tab.~\ref{tab:transfer-free} shows a comparison of the free energies we obtain from Martignac, and the reference values from the Martini 3 study. Though the hydration free energy (i.e., solvation in water) is readily calculated from Martignac, the other fields consist of transfer free energies from oil to water. These are simply computed by subtracting the two individual solvation free energies. All values are in excellent agreement of one another, within 0.3 kJ/mol for each one of them. To further demonstrate the ability to fetch the free energies from the data directly, we refer the reader to our Streamlit app, which fetches metadata from NOMAD to display the free energy resulting from each computational workflow.\cite{streamlit}

The NOMAD upload IDs for the alchemical calculations of P6 in the solvents HD, CLF, ETH, CHEX, and W are, respectively: \href{https://nomad-lab.eu/prod/v1/staging/gui/user/uploads/upload/id/PCQSjL2wQsCptzZhHal96Q}{\texttt{PCQSjL2wQsCptzZhHal96Q}}, \href{https://nomad-lab.eu/prod/v1/staging/gui/user/uploads/upload/id/GAugbmarSJmlADe8rHK4AQ}{\texttt{GAugbmarSJmlADe8rHK4AQ}}, \href{https://nomad-lab.eu/prod/v1/staging/gui/user/uploads/upload/id/HKHmUObpQ_a-SwaMVXvSeQ}{\texttt{HKHmUObpQ\_a-SwaMVXvSeQ}}, \href{https://nomad-lab.eu/prod/v1/staging/gui/user/uploads/upload/id/w4sPShVOStm9Yc-bVuFvYw}{\texttt{w4sPShVOStm9Yc-bVuFvYw}}, and \href{https://nomad-lab.eu/prod/v1/staging/gui/user/uploads/upload/id/ZwcN37wMSyidNQeQPBQbAw}{\texttt{ZwcN37wMSyidNQeQPBQbAw}}.

\subsection{Reproducibility of drug-membrane potentials of mean force}

\begin{figure}[htbp]
    \centering
    \includegraphics[width=\linewidth]{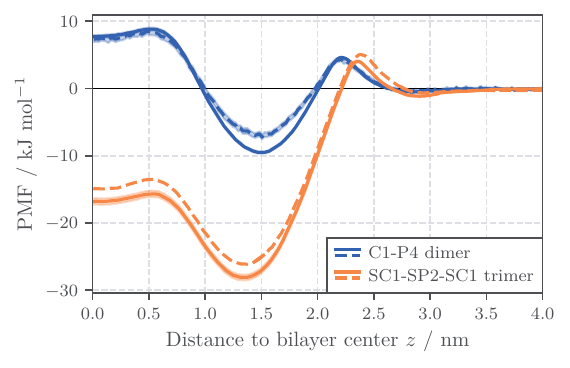}
    \caption{Drug-membrane potentials of mean force (PMFs) for a C1--P4 dimer and an SC1--SP2--SC1 trimer inserted in a POPC bilayer. Solid and dashed lines correspond to Martignac and the original studies,\citep{hoffmann2020molecular, potter2021automated}, respectively. Error estimates from bootstrapping (not available for the original trimer result) are shown as (relatively small) shaded areas.}
    \label{fig:pmf_comparison}
\end{figure}

As a second example of reproducibility, we consider the potentials of mean force (PMFs) of small Martini molecules inserted into a phospholipid bilayer. We perform PMF calculations for the C1--P4 dimer and SC1--SP2--SC1 trimer in a 1-palmitoyl-2-oleoyl-\textit{sn}-glycero-3-phosphocholine (POPC) bilayer, found originally in Hoffmann et al.\cite{hoffmann2020molecular} and Potter et al.,\cite{potter2021automated} respectively. Both studies rely on the Martini 2 force field.\cite{Marrink2003Lipids, marrink2007martini, marrink2012improved_martini2} The PMFs are calculated using the \emph{Solute-in-bilayer umbrella sampling} workflow (Fig.~\ref{fig:workflows}f). Fig.~\ref{fig:pmf_comparison} shows the original and Martignac PMFs in dashed and solid lines, respectively. The PMFs of the C1--P4 dimer show excellent agreement, with only a slight deviation around the first bead of the lipid tail region. As Hoffmann et al.~provided simulation input files, reproducing the simulations was straightforward. The update to a more recent version of GROMACS together with statistical convergence likely explain the slight variations between the PMFs. For the SC1--SP2--SC1 trimer, the Martignac PMF generally matches the result of the original study, but is slightly shifted down due to deviations in the water region ($z \gtrsim \SI{2.6}{\nano\meter}$) used as the zero reference.  Although Potter et al.~do not provide simulation input files, they include all essential parameters in their method description. We utilized their specified parameters in conjunction with defaults for the unspecified parameters. However, in contrast to extracting parameters from the method description, providing a complete input file facilitates the reproduction of a simulation and generally prevents missing parameter information. Martignac uploads all simulation input and output files pertaining to each PMF calculation as a single upload to NOMAD. As the workflow also incorporates WHAM calculations, the resulting PMF curves are included in the NOMAD upload. For the dimer and the trimer, the corresponding NOMAD upload IDs are \href{https://nomad-lab.eu/prod/v1/staging/gui/user/uploads/upload/id/VpbwP6VpS4ucuwqVBHPzeg}{\texttt{VpbwP6VpS4ucuwqVBHPzeg}} and \href{https://nomad-lab.eu/prod/v1/staging/gui/user/uploads/upload/id/N51F6fmXRl6BHpNEQNBpTQ}{\texttt{N51F6fmXRl6BHpNEQNBpTQ}}, respectively.

\begin{figure}[htbp]
    \centering
    \includegraphics[width=\linewidth]{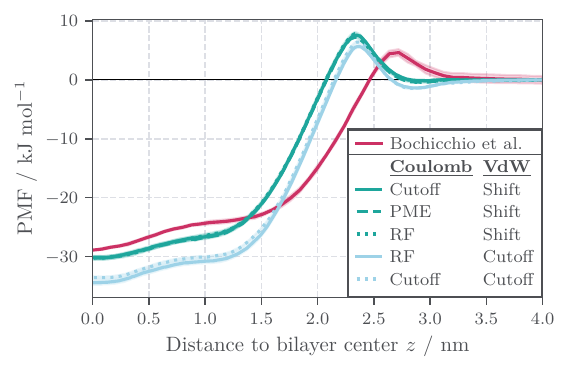}
    \caption{Comparison of multiple computations of drug-membrane potentials of mean force (PMFs) for a C1--C1 dimer inserted in a POPC bilayer with results reported by Bochicchio et al.\cite{bochicchio2015calculating} We employed various combinations of methods to address electrostatic (Coulomb) and van der Waals (VdW) interactions as implemented in GROMACS. Results using the same VdW interaction handling overlap almost entirely due to excellent agreement. We show error estimates from bootstrapping as (realtively small) shaded areas for our simulation results.}
    \label{fig:pmf_PE_comparison}
\end{figure}

We now consider the reproducibility of polyethylene, mapped to a C1--C1 dimer, in a POPC membrane from Bochicchio et al.\cite{bochicchio2015calculating} As the original study did not provide all essential simulation parameters, we could not precisely reproduce the simulation setup. While relevant parameters for the umbrella sampling, the temperature coupling, and the barostat are specified, information about the treatment of non-bonded interactions is missing. We investigate the impact of different methods for handling electrostatic and van der Waals (VdW) interactions on the resulting PMFs. Fig.~\ref{fig:pmf_PE_comparison} shows the result from the original study together with five different variants obtained with the Martignac workflow. The NOMAD upload IDs are ordered from top to bottom:
\href{https://nomad-lab.eu/prod/v1/staging/gui/user/uploads/upload/id/UkYrTZTDRUG0rTpsuJzd7Q}{\texttt{UkYrTZTDRUG0rTpsuJzd7Q}}, 
\href{https://nomad-lab.eu/prod/v1/staging/gui/user/uploads/upload/id/Eiatr72XQu6X03u0w6Tl5A}{\texttt{Eiatr72XQu6X03u0w6Tl5A}},
\href{https://nomad-lab.eu/prod/v1/staging/gui/user/uploads/upload/id/4x-JwH_IQIqqHevHf5LABQ}{\texttt{4x-JwH\_IQIqqHevHf5LABQ}},
\href{https://nomad-lab.eu/prod/v1/staging/gui/user/uploads/upload/id/owgFZkssRd6kttrlKitsTQ}{\texttt{owgFZkssRd6kttrlKitsTQ}}, and
\href{https://nomad-lab.eu/prod/v1/staging/gui/user/uploads/upload/id/XmFAOFG1Q3qd7KDBsF9mCw}{\texttt{XmFAOFG1Q3qd7KDBsF9mCw}}. While various methods for handling electrostatic interactions do not significantly impact the PMF curve, the VdW treatment causes greater differences in our results. Despite testing multiple parameters for treating non-bonded interactions, we unfortunately could not reproduce the result from Bochicchio et al. Notably, our PMF from simulations using a VdW cutoff are in excellent agreement with the C1--C1 dimer results from Hoffmann et al.\cite{hoffmann2020molecular} Additionally, our PMF more closely resembles the atomistic calculation provided as part of the original study.\cite{bochicchio2015calculating} In particular, the PMF peak near the membrane--water interface is closer to the bilayer center for their atomistic result, aligning more closely with our findings. In general, the discrepancies observed may be attributed to factors such as unsatisfactory simulation convergence or further differences in the simulation setup. For instance, the use of the polarizable water model might shift the membrane thickness, but it is unfortunately not supported by Martignac at the moment. Overall this makes a strong point for the broad and systematic use of FAIR data storage for molecular simulations. 

\begin{figure}[htbp]
  \centering
  \includegraphics[width=0.9\linewidth]{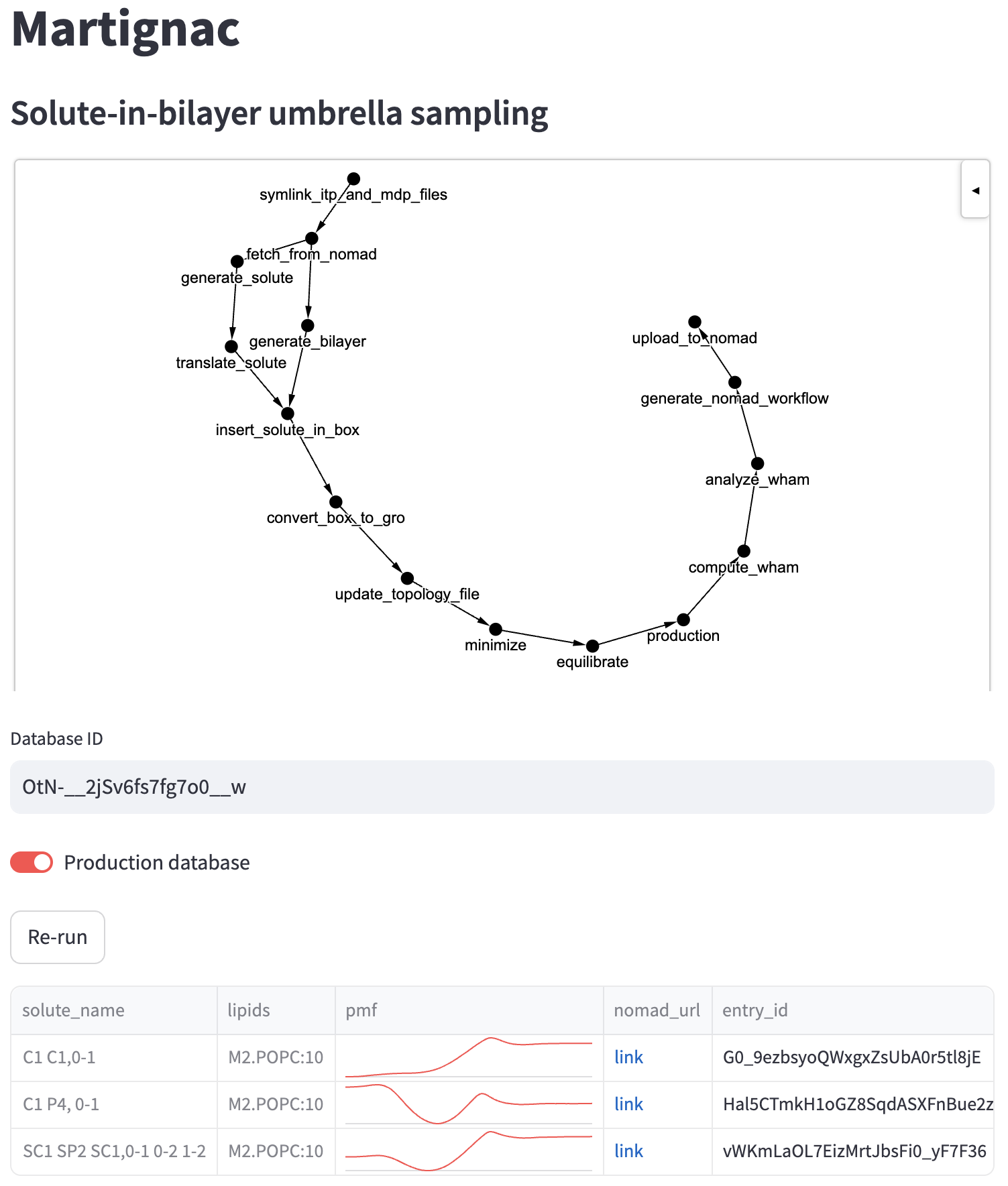}
  \caption{Snapshot of the Streamlit web app. Top: Illustration of the solute-in-bilayer umbrella sampling. Bottom: systems found in the NOMAD database \href{https://nomad-lab.eu/prod/v1/staging/gui/user/datasets/dataset/id/OtN-__2jSv6fs7fg7o0__w}{\texttt{OtN-\_\_2jSv6fs7fg7o0\_\_w}}. The query fetches the results of the WHAM calculations to automatically display the PMFs in small format.}
  \label{fig:streamlit_mini_pmfs}
\end{figure}

A notable benefit of storing simulation worfklows online is the ability to both query and display scientific information in a user-friendly fashion---here illustrated with the Martignac Streamlit app.\cite{streamlit} The app queries a NOMAD database to list all systems correponding to a Martignac workflow. In Fig.~\ref{fig:streamlit_mini_pmfs}, we show the solute-in-bilayer umbrella sampling subpage of the app. The top part of the figure displays the Martignac DAG, highlighting the branching upon system generation. Further, the bottom part shows a list of systems found in the database. For each, various information extracted from the NOMAD entry are reported. Notably, this enables us to report and display simulation outputs: here the app systematically constructs the PMF of the workflow. Such online, interactive, and visually appealing aspects are likely to promote FAIR molecular-simulation data storage.

\section{Conclusion}
\label{sec:conclusion}

We introduce Martignac: computational workflows for the coarse-grained (CG) Martini biomolecular force field. Legacy Bash scripts with error-prone copy/pasting make way for a more robust approach by means of workflows. The set of operations relevant to a particular objective (e.g., generating a box of solvent or calculating a free energy) are connected in an acyclic directed graph (DAG). The DAG links said operations to offer a \emph{traceable history} from system generation, to MD simulations, to analysis and estimation of material/thermodynamic properties. The history offers anyone the ability to inspect, check, and reproduce the content at each step. Moreover, the definition of elementary workflows enables their composability: separating system generation from further analysis means that a single instance of the former can be applied to a variety of downstream calculations. Here, we not only separate system generation from free-energy calculations, we split system generation in terms of their basic components: solutes, homogeneous liquids, and lipid bilayers. We show that Martignac greatly facilitates both reproducibility and composability by means of several examples pertaining to oil/water transfer free energies and drug-membrane thermodynamics.

The deep interconnection between Martignac and NOMAD carries interesting benefits. First, the systematic pulling of existing workflows greatly improves sustainability: the community can download existing Martini simulations, instead of simulating them (again).  The automatic \emph{pushing} of missing workflows removes any friction or efforts associated with publishing MD simulations. In this way, the user helps the community by enriching the corpus of Martini simulations available online. Finally, we find that publishing entire computational workflows offers a solution to the recent increase in the volume of scientific articles' supplementary information: all relevant data and metadata is stored and accessible in the NOMAD entries.

The connection to NOMAD also means that all simulation metadata is persistently available online. We refer the reader to the Martignac Streamlit app.\cite{streamlit} The app fetches all published Martignac simulations. The connection to the NOMAD API means that the entries are constantly updated with added simulations. Similar to MDverse, the app offers an intuitive user interface to browse through simulations. The added benefit of Martignac is the access to the simulation metadata, allowing us to automatically sort between workflows and extract scientifically meaningful information, such as free energies. Looking ahead, the incorporation of workflows for more biomolecular simulations of interest is straightforward, and will further help move the field to more systematic practices.

\section*{Acknowledgments}
We thank Brandon Butler and Corwin Kerr for discussion about the \texttt{signac} workflow library. T.B.~acknowledges support by the Deutsche Forschungsgemeinschaft (DFG, German Research Foundation) under Germany's Excellence Strategy EXC 2181/1 - 390900948 (the Heidelberg STRUCTURES Excellence Cluster). J.F.R.'s contribution was funded by the NFDI consortium FAIRmat - Deutsche Forschungsgemeinschaft (DFG) - Project 460197019. Icons on figure \ref{fig:nomad-intro} made by Freepik and Haca Studio from \url{www.flaticon.com}. 

\bibliography{biblio}

\end{document}